\documentclass[]{elsart3}

\usepackage{amssymb}
\usepackage{graphicx}

\usepackage{epsfig}

\begin{document}

\begin{frontmatter}

\title{Stochastic approximation to the specific response of the immune system}

\author{Nuris Figueroa Morales}
\address{Department of Theoretical Physics, Physics Faculty, University of Havana, La Habana, CP 10400, Cuba}
\address{``Henri-Poincar\'e-Group'' of Complex Systems, Physics Faculty, University of Havana, La Habana, CP 10400, Cuba}
\author{Kalet Le\'on}
\address{Center for Molecular Immunology, P.O. Box 16040, Habana CP 11600, Cuba}
\author{Roberto Mulet}
\address{Department of Theoretical Physics, Physics Faculty, University of Havana, La Habana, CP 10400, Cuba}
\address{``Henri-Poincar\'e-Group'' of Complex Systems, Physics Faculty, University of Havana, La Habana, CP 10400, Cuba}
\date{September 2010}

\begin{abstract}

We develop a stochastic model to study the specific response of the immune system. The model is based on the dynamical interaction between  Regulatory and 
Effector CD4+ T cells in the presence of Antigen Presenting Cells inside a lymphatic node. At a mean field level  the model predicts the existence of different regimes  where active, tolerant, or cyclic immune responses are possible. To study the model beyond mean field and to understand the specific responses of the immune system we use the Linear Noise Approximation and show that fluctuations due to finite size effects may strongly alter the mean field scenario. Moreover, it was found the existence of a certain characteristic frequency for the fluctuations. All the analytical predictions were compared with simulations using the Gillespie's algorithm.

\end{abstract}

\end{frontmatter}


\maketitle

\section{Introduction}
\label{sec:int}

The purpose of the immune system is to detect and neutralize the molecules, or cells, dangerous for the body without damaging the healthy cells. 
A fundamental process of the immune system is the maintenance of self-tolerance, i.e., the prevention of harmful immune responses against body components \cite{Langman}. The biological significance of this process becomes very patent upon its failure during pathological conditions known as
autoimmune diseases. 

The risk of autoimmunity cannot be dissociated from the capacity of the immune system to cope with diverse and fast evolving
pathogens. The latter is achieved by setting up a vast and diverse repertoire of antigen receptors expressed by lymphocytes, which as a
whole is capable of recognizing any possible antigen. Most lymphocytes have a unique antigen receptor (immunoglobulin in B-cells and TCR in T
cells) that is encoded by a gene that results from somatic mutation and random assortment of gene segments in lymphocyte precursors. The
randomness in the generation of antigen receptors makes it unavoidable that lymphocytes with receptors recognizing body antigens are also made.
These autoreactive lymphocytes can potentially cause autoimmune diseases if their activation and clonal expansion are not prevented. The question
is how is this avoided in healthy individuals?

Regulatory CD4+ T cells, which express forkhead box protein 3 (FoxP3)  are enriched in the CD25 pool of healthy individuals  have been
gaining increasing relevance in immunology \cite{Sakaguchi}. Many lines of evidence indicate that these cells play a key role in the development of natural
tolerance and in the prevention of autoimmune pathologies, by controlling the activation and proliferation of other autoreactive
lymphocytes. The functional significance of these cells has broadened, as they were shown to modulate the immune response against pathogens,
preventing the associated immunopathology , and the rejection of transplants.

Some years ago, Le\'on et al \cite{Kalet} proposed a simple model, consistent with these results able to explain the mechanism of immunological self-tolerance.
The main hypothesis behind this model is that regulatory cells inhibit the proliferation of effector T cells, but depend on the latter for their own proliferation. The interaction between these cells is mediated by the presence of Antigen Presenting Cells (APC).

The model, was developed to study the response of macroscopic
populations of lymphocytes, i.e.,the $10^7$ lymphocytes that may be
present in a single lymph node and was technically approached
studying the stability or numerically solving the corresponding
differential equations \cite{Kalet}. This approach however, might be
inappropriate to study the specific clonal responses in the immune
system.

The $10^7$ lymphocytes that exist in a lymph node, are really divided in
approximately $10^5$ or $10^6$ different clones\cite{Perelson}. I.e. sets of lymphocytes which recognize the same antigen and participate in their own specific,
rather independent, immunological response. Therefore, the number of
cells involved in any particular clonal response is far smaller: $10^2$ or
$10^4$ T cells, which interact specifically with a small subset of all the
antigen presenting cells available in the lymph node. One must naturally
expect to found clones of different sizes, perhaps reflecting the
ubiquitous nature of the antigen recognized by the T cells (i.e. the
amount of APCs which are recognized). However, the small size expected
for most T cells clones would made differential equations an
inappropriate tool to describe their dynamics.

In this work, we first reformulated the model of Kalet and co. \cite{Kalet} within a more general stochastic frameworks of  interactions between regulatory (R) and effector (E) T-cells.  From this formulation we derive, on one hand a mean field description
 similar -but not identical- to the ones already known in the literature  for the response of the {\em entire} immune system \cite{Carneiro}. On the other, a self-consistent approach to unveil the role of the fluctuations in the {\em specific} response of the system.

The rest of the paper is organized as follows. In the next section, that may be bypassed in a first reading of the manuscript, we present the main mathematical techniques involved in our work. Section \ref{sec:mod} introduces and motivates the model. In section \ref{sec:res} we first compare, at the mean-field level, the predictions of our model with those of similar approaches in the literature. Then, we discuss
how fluctuations may affect these predictions highlighting  their relevance for the clonal response of the 
immune system. Finally, the conclusions and some remarks suggesting possible extensions of our work appear in Section \ref{sec:conc}.

\section{Mathematical Techniques}
\label{sec:mat}

When the effects of fluctuations are relevant, as it is often the case in biological systems, a modeling 
approach based on ordinary differential equations is not appropriate.

This is well understood when the source of noise are the thermal fluctuations. But, while less mentioned, it is also important when the noise
is induced by the finite number of elements in the populations, i.e., molecules, cells, or individuals \cite{ElfEhren}.

A more convenient framework to describe these problems is a stochastic approach taking into account the probabilistic nature of every process. The natural procedure is to write down a differential-difference equation that keeps
track in time of the probability that the system would have a certain configuration ($\vec{n}$). Under the Markov assumption it is called the Master equation \cite{Reichl}

\begin{eqnarray}
 \frac{\delta P(\vec{n},t)}{\delta t} = \sum_{\vec{n'} \neq \vec{n}} P(\vec{n'},t) W_{\vec{n'}\rightarrow \vec{n}} - \nonumber \\
 \sum_{\vec{n'} \neq \vec{n}}P(\vec{n},t) W_{\vec{n} \rightarrow \vec{n'}}
\label{eqnM} 
\end{eqnarray}
where $\vec{n}$ is a vector whose components are the numbers of elements of every different
 species in the system, $P(\vec{n},t)$ is the probability of having the state $\vec{n}$ at time $t$, and 
$W_{\vec{n'}\rightarrow \vec{n}}$ is the probability of the transition $\vec{n'}\rightarrow \vec{n}$ per unit time.

Unfortunately, the Master equation can only be solved in special cases, usually, when the transition probabilities are linear or constant in the state variables \cite{Gardiner}. In more general situations we must, instead, resort to approximation methods.  These approximation methods may be divided into two  broad classes: numerical simulation algorithms, and perturbative calculations. Under the first class the Gillespie \cite{Gillespie} algorithm has become the standard tool in the community  (see appendix for a short description of the algorithm). It is a Monte Carlo's algorithm that simulates a single trajectory $\vec{n}(t)$ consistent with the unknown probability distribution
$P(\vec{n},t)$ characterized by the Master equation. It guarantees to find the correct solution of the Master equation, but it is often time consuming and it is not suitable for parameter exploration. On the contrary, perturbative methods lack the confidence given by a full and formal solution of the Master equation, but  may provide important insight on the behavior of the system and on the relevance of the parameters. 

In this work we present two different perturbative solutions. The {\em  Linear Noise Approximation} ( LNA) \cite{vanKampenlibro} which is an expansion around the large size solution of the corresponding Master equation. This gives us a tool to understand the relative size of the fluctuations and their relations with the parameters of the model. On the other hand, the 
{\em Effective Stability Approximation} ( ESA) \cite{Scott}, which is an extension of the LNA,  provides a quantitative characterization
 of the effect of the  fluctuations in the stability of the phases predicted by the corresponding mean field model.

\subsection{The Linear Noise Approximation}
\label{subsec:LNA}

An important difficulty in solving the Master equation arises from the discreteness of the 
variables involved ($\vec{n}$). As long as the number of elements increases the system 
evolution turns more regular and the mean-field equations give a more accurate description.
 The Linear Noise Approximation is a systematic approximation method that rests upon
 the assumption that the deterministic evolution of the concentrations in the system can be 
meaningfully separated from the fluctuations, and that these fluctuations scale roughly as 
the square root of the system size $\Omega$.
 
Moreover, in systems where the size of the populations differs in orders of magnitudes, it is expected that the respective concentrations fluctuate within different scales. For this reason, we consider as many parameters $\Omega_i$ as species are involved in the reactions, and write the numbers of individuals per species proportional to these values. Each $\Omega_i$ will be a characteristic size scale for every particular specie.

Therefore, under the LNA, the population numbers per specie are written as:
\begin{equation}
 n_i = \Omega_i x_i + \sqrt{\Omega_i} \alpha_i\mbox{,}
\label{lna}
\end{equation}
where $x_i$ is the deterministic prediction for the concentration of the $i^{th}$ species 
with respect to the parameter $\Omega_i$, and $\alpha_i$ measures the fluctuations around $x_i$.  Note that formalizing the LNA in this way, there isn't any {\itshape a priory} assumption about the system size and concentrations higher than one are allowed. If for every specie $\Omega_i=\Omega_j$ the standard formulation is recovered.

Now we shall use the continuous variables $\alpha_i$ instead of the integers $n_i$ to write the probability distribution $P(\vec{n},t)$. Let us group the magnitudes $x_i$ into the macroscopic concentrations vector 
$\vec{x}=(x_1,x_2,\ldots,x_N)$, and the $\alpha_i$ into the fluctuations vector $\vec{\alpha}=(\alpha_1,\alpha_2,\ldots,\alpha_N) $.
The matrix $\Omega= diag(\Omega_1,\Omega_2,\ldots,\Omega_N)$ is a diagonal matrix whose principal elements are the extensive parameters $\Omega_i$
  with the units of the volume to which concentrations $x_i$ are refered.

Let us define the following variables:

\begin{eqnarray}
\sigma_i=\frac{\Omega_1}{\Omega_i}  \mbox{,} \;\;\;\  \omega=\frac{1}{\sqrt{\Omega_1}}     \;\;\;\ \mbox{and} \;\;\;\    \beta_i=\sqrt{\sigma_i}\alpha_i \mbox{,}  
\label{variables1}
\end{eqnarray}
thus, we have
\begin{equation}
 n_i=\Omega_i x_i + \Omega_i \omega \beta_i \;\ \mbox{.}
 \label{variables2}              
\end{equation}
Taking (\ref{variables2}) into account to expand the Master equation (\ref{eqnM}) in powers of $\Omega_i$, at the leading order  the following system of differential equations appears: 

\begin{equation}
 \frac{\partial x_i}{\partial \tau} =  \sigma_i \sum_{j=1}^{M}  S_{ij}  \tilde{W}_j(\Omega \cdotp \vec{x}) \mbox{,}
\label{determinista}
\end{equation}
This corresponds to the  deterministic rate equations that are often used to describe the dynamics of such 
systems at a mean-field level.

 Here the time scale has been modified following $t=\Omega_1 \tau$ and $S_{ij}$ is the stoichiometry coefficient, the number in what the $i^{th}$ species 
varies when the $j^{th}$ process occurs. These coefficients will correspond with the element of row $i$ and column $j$ of the stoichiometry matrix $\textbf{S}$.

In the next order approximation one finds:

\begin{eqnarray}
 \frac{\partial \Pi(\vec{\beta},\Omega_1\tau) }{\partial \tau} = -\sum_{i,k=1}^{N} A_{ik} \frac{\partial[\Pi(\vec{\beta},\Omega_1\tau) \beta_k]}{\partial \beta_i} + \nonumber \\ 
\frac{1}{2} \sum_{i,k=1}^{N} D_{ik} \frac{\partial^2 \Pi(\vec{\beta},\Omega_1\tau)}{\partial \beta_i \partial \beta_k}   \mbox{,}
\label{FokkerPlank}
\end{eqnarray} 
where
\begin{equation}
 A_{ik}  = \sigma_i \sum_{j=1}^{M} S_{ij} \frac{\partial \tilde{W}_j(\Omega \cdotp \vec{x})}{\partial x_k}  
\label{A}
\end{equation}
and
\begin{equation} 
 D_{ik}=\sigma_i \sigma_k\sum_{j=1}^{M} S_{ij} S_{kj} \tilde{W}_j(\Omega \cdotp \vec{x})\mbox{.}
\label{D}
\end{equation}
$A_{ik}$ and $D_{ik}$ are respectively the elements ($i,k$) of the matrices $\textbf{A}$ and 
$\textbf{D}$, where, 
$\textbf{A}$ can be identified with the Jacobian matrix of the system of differential equations
(\ref{determinista}).

Note that equation (\ref{FokkerPlank}) is a Fokker-Plank equation that characterizes the probability 
distribution for the fluctuations  $\Pi(\vec{\beta},\Omega_1\tau)$, centered on the macroscopic
trajectory $\Omega \cdotp \vec{x}(t)$.

The matrices $\textbf{A}$ and $\textbf{D}$ are independent of $\vec{\beta}$, which appears 
only linearly in the drift term. As a consequence, the distribution $\Pi(\vec{\beta},t)$ will
be Gaussian for all time \cite{Risken}. In particular, at equilibrium ($\vec{x}=\vec{x}_s$) 
the fluctuations are distributed with density 
\begin{equation}
 \Pi_s (\vec{\beta})= \left[ (2 \pi)^N \textit{det}[\Xi] \right]^{-\frac{1}{2}} \textit{exp} \left[-\frac{1}{2} \vec{\beta}^T \Xi \vec{\beta}\right] \mbox{,} 
\nonumber
\end{equation}  
and variance $\Xi=\left\langle \vec{\beta} \cdot \vec{\beta}^T\right\rangle $ determined by

\begin{equation}
 \textbf{A} \cdot \Xi + \Xi \cdot \textbf{A}^T + \textbf{D}=0 \mbox{,}
\label{eqnmatricial}
\end{equation} 
where $\textbf{A}$ and $\textbf{D}$ are evaluated on the studied fixed point \cite{Risken}.

The steady-state time correlation function is  
\begin{equation}
\left\langle \vec{\beta}(\tau) \cdot \vec{\beta}^T( \tau-\tau')\right\rangle = \textit{exp}\left[\textbf{A} \tau' \right] \cdot \Xi  \mbox{.} 
\label{relacion}
\end{equation} 

Moreover, if fluctuations are important it may be useful to study their properties. With this aim, a very powerful tool is the Fourier analysis of the equations that govern it. In our case it is the Fokker-Plank equation
(\ref{FokkerPlank}), who gives us all the information about the temporal behavior of fluctuations. Given the
 inconvenient form of this equation for our purpose, it is more reasonable to write it in a completely equivalent formulation more benevolent to investigation using Fourier transforms \cite{McKane}.   
The problem can then be formulated as the set of stochastic differential equations of the Langevin type \cite{Risken}:   

\begin{equation}
 \frac{d \beta_i}{d \tau} =\sum_{k=1}^N A_{ik} \beta_k  + \eta_i(\tau)\mbox{,}
\label{Langevin}
\end{equation}
where $\eta_i(\tau)$ is a Gaussian noise with zero mean and correlation function given by
\begin{equation}
 \left\langle \eta_i(\tau) \eta_k(\tau') \right\rangle = D_{ik} \delta (\tau-\tau')\mbox{.}
 \label{correlacion}
\end{equation}

The power spectrum of the fluctuations can be found for every specie of the system by averaging the square modulus of the
 Fourier transformation of  $\beta_i$. In this way,
\begin{eqnarray}
 P_i(\omega) & = & \left\langle |\tilde{\beta}_i(\omega)|^2 \right\rangle \nonumber \\ 
 & = &  \sum_{j=1}^N \sum_{k=1}^N \Phi^{-1}_{ij}(\omega) D_{jk} (\Phi^\dag)^{-1}_{ki}(\omega) \mbox{.}
\label{espectropotencias1}
\end{eqnarray}
Being $\Phi_{ik} = i \omega \delta_{ik} - A_{ik}$ and $\Phi^\dag_{ki}(\omega)=\Phi_{ik}(-\omega)$ \cite{McKane}.
 With this expression we have an analytic measure of the contribution of every frequency to the oscillatory  behavior of the concentrations around each fixed point. Through it, the properties of the fluctuations can be studied.

\subsection{Effective Stability Approximation}
\label{sec:ESA}

It is required just a bit of intuition to realize that the  noise can affect 
the predictions of the deterministic rate equations models \cite{Scott}. This is particularly clear, once we question about the stability of the fixed point predicted by a mean-field model.

In the deterministic analysis, the stability of the model $\frac{d \vec{x}}{d \tau} = \vec{f}(\vec{x})$ to small perturbations
is found by linearizing about the equilibrium point: $\vec{x}=\vec{x}_s + \vec{x}_p$ and the eigenvalues of the Jacobian $\textbf{J}^{(0)}=\left. \frac{\partial \vec{f}}{\partial \vec{x}}\right|_{\vec{x}=\vec{x}_{s}}$ provide the decay rate of the exponential eigenmodes \cite{Strogatz}.

If we now consider the fluctuations around the small perturbation $\vec{x}_p$, 
the concentrations remain $\vec{x}=\vec{x}_{s} + \vec{x}_p + \omega \vec{\beta}(\tau)$. The Jacobian matrix
 $\textbf{J}=\left. \frac{\partial \vec{f}}{\partial \vec{x}}\right|_{\vec{x}=\vec{x}_{s}+\omega \vec{\beta}(\tau)}$, 
will be  expressed in terms of the fluctuations $\omega \vec{\beta}(\tau)$ around the steady-state.
In the limit $\omega \longrightarrow 0$, we can linearize $\textbf{J}$ with respect to $\omega$ and the new Jacobian becomes

\begin{eqnarray}
 \textbf{J} \approx \left. \textbf{J} \right|_{\omega \rightarrow 0} +  \left.\omega \frac{\partial \textbf{J}}{\partial \omega} \right|_{\omega \rightarrow 0}  \equiv \textbf{J}^{(0)} + \omega \textbf{J}^{(1)}(\tau)  \mbox{,}
\label{jacobianocorregido}
\end{eqnarray} 

Therefore, new {\em effective} eigenvalues ${\lambda'}_i$, including the influence of the intrinsic noise, characterize the stability of the phases \cite{Scott}.

\begin{equation}
 {\lambda'}_i=\lambda_i + \lambda_{i_{corr}}\mbox{,}
\label{lambdacorregida}
\end{equation}

\noindent where $\lambda_{i_{corr}}$ is proportional to $\omega$, i.e., inversely proportional to the system size.

When the system size is small, fluctuations are relevant, and in many cases the corrections $\lambda_{i_{corr}}$ can 
not be ignored. This small correction may, indeed, change the sign of the eigenvalues (now ${\lambda'}_i$) and as a consequence, to change the stability of the phase.

\section{The cross-regulatory model}
\label{sec:mod}

In this section we motivate and present a stochastic model to describe the dynamics of two (possible small) populations of T cells E and R of the immune system. 

Some clues on how the regulatory CD4+ T cells suppress the response of other cells have been derived from well-correlated in vitro and in vivo experiments. These studies suggest that T-cell-mediated suppression is not mediated by soluble factors and require cell-to-cell contact \cite{Takahashi}\cite{Thornton}. Moreover, regulatory CD4+ T cells can only suppress the response of other cells if the ligands of both cells are expressed by the same Antigen Presenting Cell (APC).
It also has been seen that regulatory T cells do not proliferate in vitro when cultivated alone in the presence of APCs.

\begin{figure}[!htb]
     \begin{center}
                 \includegraphics[angle=0, scale=0.20]{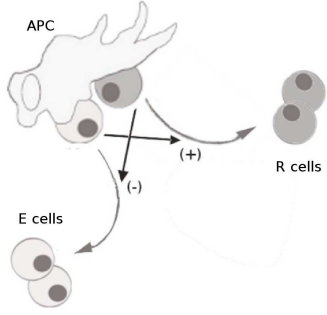} 
     \end{center}
 \caption{Cartoon illustrating the hypothetical mechanism presence-proliferation during simultaneous conjugation 
between $T$ lymphocytes and APCs. E cells promote R cells 
proliferation, while R cells avoid E proliferation.}
  \label{fig:figurita}
\end{figure}

Despite these strong requirements, it is not yet clear what the nature of this mechanism is.
The model requires a set of postulates that summarize the life cycle of these cells, and 
their interactions. Inspired by the success of  Le\'on et al.\cite{Kalet} we  are going to 
assume that the proliferation of T cells occurs via its conjugation with an APC. 
An E (effector) cell could duplicate only if conjugated with an APC which has no R (regulatory) cell.
On the other hand, an R cell can duplicate only if conjugated with an APC where at least one E cell is also conjugated. In that way, regulatory T cells exist only in the presence of effector T cells, but the growth of the 
E cells is regulated by the presence of the R cells. A picture illustrative of these
postulates is shown in Figure \ref{fig:figurita}.


In vivo the total number of APCs and their capacities to stimulate T cells may change, and it is
 perhaps a function of the T cells themselves, but we consider them as a constant population
in which every APC has a finite and fixed number of conjugation sites.
Then, each site can be empty or occupied by a regulatory or effector T cell and all the cells that belong to the same phenotype (E or R) will have the same probability of being conjugated with any free site of any APC.

Within these postulates we have to deal with the populations of free and conjugated cells for each specie, and the transition between the different states. This makes the model very cumbersome at the stochastic level, and almost intractable.   Fortunately the processes of formation and dissociation of the conjugates are relatively fast compared to the overall dynamics of the T cell populations. While a T cell remains conjugated with its APC for a few hours, the T cell mitotic cycle lasts on average 12 hours and the T cell lifespan is at least several days. Then, it is reasonable to assume that only free lymphocytes die, and that the numbers of conjugated T cells of  both phenotypes are in quasi-steady state.

Therefore, if  $a$ is the total number of APCs, and $s$, the number of conjugation sites per APC we may define:

\medskip
$ k_{on_E} E \left( \frac{a s-E^b-R^b }{a s}\right) \equiv$ Probability per unit time of combination of 
an E cell and an APC.

\medskip
$ k_{off_E} \frac{E^b}{a s} \equiv$ Probability per unit time of dissociation of E to its conjugate.

\medskip
Here $E$ and $E^b$ are the numbers of free and conjugated E cells respectively, and  $ k_{on_E}$ and $k_{off_E}$ are positive parameters that characterize the processes of occurrence and disappearance of the conjugates of this cellular species.

In the same way, for R cells these probabilities per unit time are given by
 $k_{on_R} R \left( \frac{a s-E^b-R^b }{a s}\right)$ and $k_{off_R}\frac{R^b}{a s} $ respectively.

If now we set $q_e= \frac{k_{on_E}}{k_{off_E}}$ and $q_r= \frac{k_{on_R}}{k_{off_R}}$, the stationary state for the formation and dissociation of conjugates for every species reads:

\begin{eqnarray}
 E^b= \frac{a s q_e E}{1+q_e E+q_r R}  \;\;\;\   R^b= \frac{a s q_r R}{1+q_e E+q_r R} \mbox{.}
\label{EyRpegadas}
\end{eqnarray}

\begin{figure}[!htb]
     \begin{center}
                 \includegraphics[angle=0, scale=0.30]{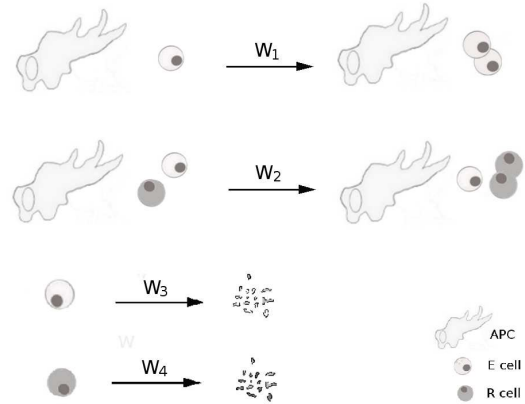}
     \end{center}
 \caption{Reaction diagram indicating the events underlying the dynamics of APCs, E cells and R cells, as assumed in the model.}
  \label{fig:4procesos}
\end{figure}

After this approximation, our analysis involves only four significant processes, the birth and death of effector and regulatory cells, and is far more tractable.
The four processes are represented in Figure \ref{fig:4procesos}.

Each event is characterized by an occurrence probability per unit time  $W(\vec{n} \rightarrow \vec{m})$ from the state $\vec{n}=(E,R)$ to the posterior one $\vec{m}=(E',R')$.
It is,

 $$
W(\vec{n} \rightarrow \vec{m}) =
   \left\{
       \begin{array}{lcr}
           W_1 & & \mbox{if}\;\ \vec{m}=(E+1,R) \\
           W_2 & & \mbox{if}\;\ \vec{m}=(E,R+1)  \\
           W_3 & & \mbox{if}\;\ \vec{m}=(E-1,R)  \\
           W_4 & & \mbox{if}\;\ \vec{m}=(E,R-1)  
       \end{array}
    \right.
$$
and the  stoichiometry matrix corresponding to this system, whose element (\textit{i},\textit{j}) contains the changes produced on the specie $i^{th}$ when it occurs the $j^{th}$ process is

\begin{equation}
\textbf{S}= \left( \begin{array}{*{4}{r}}
1 & 0 & -1 & 0  \\
0 & 1 & 0 & -1
\end{array}   \right)\mbox{.} 
\label{S}
\end{equation}


 Now, following the postulates of our model, a new E cell appears in the system
 if there already exists a combined E cell with an APC that has no R cells simultaneously
 conjugated. Then, the probability of the increase of $E$ in $1$ is proportional to the number of APCs ($a$),
times the probability to find in it a conjugated E cell ($\frac{E^b}{as}$), times the probability that this APC has no R cells simultaneously combined. The later has the form of an Hypergeometric distribution
\cite{Kalet}, but
 to gain in simplicity of the expressions we can manage it approximately as $\left( 1-\frac{R^b}{as}\right)^{s-1} $,
 valid when $a \geq 10$ \cite{Karina}. This takes us to the expression 
\begin{equation}
W_1=\psi_e a\frac{E^b}{a s}\left( 1-\frac{R^b}{a s} \right)^{s-1} \mbox{.}
\label{lambdae}
\end{equation} 

The analysis for the growth  of the R population  in $1$ unit is similar. 
A new regulatory cell appears in the system
 with certain probability if there exists already both R and E cells conjugated in the same APC during the same interval of time. 
The probability of the existence of an E cell conjugated in a given APC where in addition exists space for an R cell can be written as $\left[1-\left(1-\frac{E^b}{as} \right)^{s-1}  \right]$. Multiplying 
 this by the probability of randomly find that specific APC with an R cell combined in it gives the probability for this birth:

\begin{equation}
W_2=\psi_r a\frac{R^b}{a s}\left[ 1- \left( 1-\frac{E^b}{a s} \right)^{s-1} \right]\mbox{.}
\label{lambdar}
\end{equation} 
In equations (\ref{lambdae}) and (\ref{lambdar}), $\psi_e$ and $\psi_r$ are parameters for the effector and regulatory species that characterize the processes of formation and dissociation of the different conjugates. 

Finally, we assume that the death probabilities for each specie are proportional to the numbers of  individuals, with proportionality factors $m_e$ for effector and $m_r$ for regulatory.

Let us set $e$ the concentration of E cells with respect to the quantity $\Omega_e$, and $r$ will be the 
concentration of R cells relative to $\Omega_r$, where $\Omega_e$ and $\Omega_r$ are the characteristic mean values. The probability vector per unit time $\vec{W}$, that contains the information of the four processes of birth and death is formed by the elements:

\begin{eqnarray}
 W_1&=&\frac{\psi_e a q_e e \Omega_e}{1+q_e e \Omega_e}\left(\frac{1+ q_e e \Omega_e}{1+ q_e e \Omega_e+ q_r r \Omega_r}\right)^s  \label{W1} \\
 W_2&=&\frac{\psi_r a q_r r \Omega_r}{1+ q_e e \Omega_e+ q_r r \Omega_r} \times  \nonumber \\
   &&\left( 1-\left( \frac{1+q_r r \Omega_r}{1+ q_e e \Omega_e+ q_r r \Omega_r}\right)^{s-1}\right)  \label{W2}\\
 W_3&=&m_e e \Omega_e \\
 W_4&=&m_r r \Omega_r\mbox{.}   \label{W4}
\end{eqnarray}  
where the dependencies with $E^b$ and $R^b$ that appear in (\ref{lambdae}) and (\ref{lambdar}) have been substituted by the expressions in (\ref{EyRpegadas}).


\section{Results and Discussion}
\label{sec:res}

\subsection{Mean-field approximation}

Once we know all the elements of the transition probabilities per unit time, and the stoichiometry of the events, we are in conditions to write the deterministic rate equations that correspond to our model (see eq. (\ref{determinista}) ).

\begin{eqnarray}
 \frac{d e}{d \tau}&=&W_1-W_3  \label{ecpuntofijo} \label{det1}\\ 
 \frac{d r}{d \tau}&=&\frac{\Omega_e}{\Omega_r}W_2- \frac{\Omega_e}{\Omega_r}W_4 \mbox{,}
\label{det2}
\end{eqnarray}
where the time has been measured according to $t=\Omega_e \tau$.

In the steady-state, $\frac{d e}{d \tau}= \frac{d r}{d \tau}=0$, the solutions for the deterministic system are given by a fixed pair ($e^{\ast},r^{\ast}$) of numbers. Depending on the parameters, the stationary
 solutions with biological sense may be of three types: ($0,0$), ($e^{\ast}>0,0$) and ($e^{\ast}>0,r^{\ast}>0$).

\begin{figure}[!htb]
     \begin{center}
                 \includegraphics[angle=0, scale=0.4]{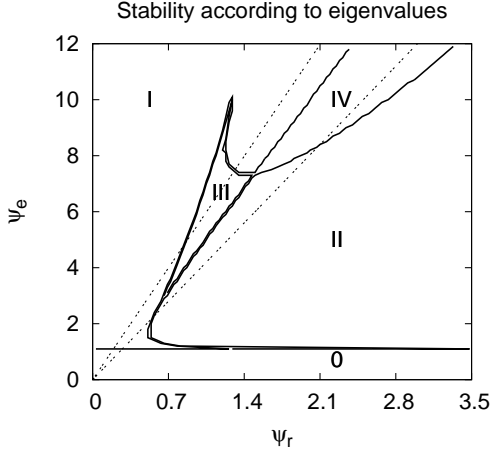}
     \end{center}
 \caption{Keeping constant the parameters $s=3$, $a=1000$, $\Omega_e=1000$, $\Omega_r=2000$, $q_e=10^{-3}$,
 $q_r=5\cdot10^{-4}$, $m_e=1.0$ and $m_r=0.1$, the five parameter regions with 
different kinds of stability depending on the [$\psi_r$;$\psi_e$] pairs are shown.
Each ordered pair was classified attending to the signs of the eigenvalues of the Jacobian evaluated 
on the fixed points. 
 Solid lines represent the region borders.}
  \label{fig:mapa}
\end{figure}

Attending to the stability of these solutions we define five phases that are sketched in  Figure \ref{fig:mapa}.

{\bf Phase 0}: In this phase only the trivial fixed point ($0,0$) exists. It is characterized by the relation $ \psi_e a q_e <m_e$ and within this phase, this fixed point is always stable.

{\bf Phase I}: In this phase, only the point ($e^{\ast}>0,0$) is stable.

{\bf Phase II}: Only the point ($e^{\ast}>0,r^{\ast}>0$) is stable.

{\bf Phase III}: It is a bistable phase. Both points ($e^{\ast}>0,0$) and ($e^{\ast}>0,r^{\ast}>0$) are stable

{\bf Phase IV}: In this phase, the fixed points presented above are unstable. A limit cycle develops.

\medskip

This phase diagram is very similar to the one reported in \cite{Kalet}. However, for completeness and because the Phase IV was not predicted before we discuss in detail below the biological interpretation of the five regions.

 Since E cells themselves do not mediate the effective immune response, but they trigger it, we interpret the state dominated by regulatory T cells ($e^{\ast},r^{\ast}$) as tolerance or unresponsiveness, and the state 
dominated by effector ($e^{\ast}>0,0$) as an effective immune response or autoimmunity. Such interpretation arises from the fact that when R cells exist in the system, this population competes with the effectors for the APCs, limiting the growth of the E population.

Particularly relevant from the immunological point of view is Phase III. Operating in this region, the model can  switch between the states dominated by regulatory or effector T cells in a reversible way \cite{Kalet}.


For the fixed points of the kind ($e^{\ast}>0,0$), $e^{\ast}=\frac{\psi_e a q_e - m_e}{m_e q_e \Omega_e}$, the eigenvalues of the Jacobian matrix are:
\begin{equation}
 \lambda_{1(e^{\ast},0)}= m_e \Omega_e \left(-1 + \frac{me}{a q_e \psi_e}\right)
\label{negativo}
\end{equation}
and
\begin{eqnarray}
 \lambda_{2(e^{\ast},0)} & = & -\Omega_e m_r + \nonumber \\
&  & \Omega_e q_r \psi_r\left(\frac{m_e}{q_e \psi_e} -a \left(\frac{m_e}{a q_e \psi_e}\right)^s   \right) \mbox{,} 
\label{positivo}
\end{eqnarray} 
the first one of these eigenvalues is always negative and its corresponding 
eigenvector is $v_{1(e^{\ast},0)}=(1,0)$. It guarantees that, if not perturbed along the $r$ direction,  the fixed point is stable. This means that in those phases where both species can coexist ($II$, $III$ and $IV$), once the R population is extinguished,  to take the system away from the ($e^{\ast},0$) state it is necessary that at least one R cell comes from outside the system. The absence of such a perturbation can be interpreted as the lost of tolerance, and the appearance of an effective immune response. It explains the fact that thymectomized animals are  more susceptible to procedures that induce autoimmunity that euthymic animals \cite{Shevach}.

It is also well known that if in a tolerant organism the number of APCs is increased enough, the immune system may turn on an immunological response. To study under which conditions our model reproduces this effect it is useful to understand that to change the parameters space from a state given by ($a,\psi_r,\psi_e$) to another characterized by  ($na,\psi_r,\psi_e$) is equivalent, in terms of the deterministic equations, to change the state ($a,\psi_r,\psi_e$) to ($a,n \psi_r,n \psi_e$) along the straight line that matches  them. So, the increase in the number of APCs ($a$) may be interpreted as a motion along a straight line with slope $\frac{\psi_e}{\psi_r}$ that passes through the origin in the $\psi_r$ \textit{vs} $\psi_e$ map.

In our model, a tolerant organism is located in phases $II$ or $III$. In the case of a system that starts in region $III$, raising the number of APCs makes the system first, to evolve into region $I$ (see the line with points in  Figure \ref{fig:mapa}). Biologically, it may be interpreted as the entrance into a region where the tolerance disappears and the animals  develops immunological response. On the other hand, when the organism starts in the phase $II$, increasing the number of APCs drifts the system into a cyclic immune response (see the dashed line in  Figure \ref{fig:mapa}).

\begin{figure}[!htb]
     \begin{center}
                 \includegraphics[angle=0, scale=0.4]{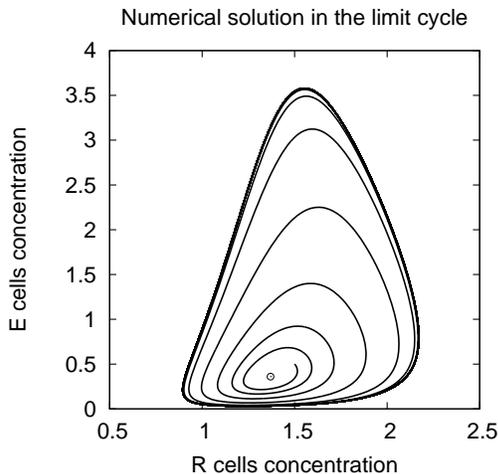} 
     \end{center}
 \caption{With the parameters of Figure \ref{fig:mapa}, and the pair $[\psi_r=2.2$; $\psi_e=11.0]$, it is shown the numerical solution for the deterministic rate equations (\ref{det1}) and (\ref{det2}). The trajectory surrounds the unstable fixed point ($e^{\ast},r^{\ast}$)(represented by a white circle) and evolves to a cyclic behavior.}
  \label{fig:ciclo}
\end{figure}

It is worth to highlight that our mean-field predictions compare very well with those obtained by K. Le\'on \textit{et al.}\cite{Kalet} for a similar model of R-E interactions regulated by APCs. However, our phase diagram is a bit richer. Exploring a parameter region where $q_e \neq q_r$ the bi-stable phase III may be  now bounded by a phase $IV$ where the biologically meaningful fixed points are unstable. Using parameters in this zone we have numerically solved the system of differential equations (\ref{det1}) and (\ref{det2}). As Figure \ref{fig:ciclo} shows, in this zone appear very well defined closed orbits, representative of limit cycles.

The existence of this limit cycle allows the model to reproduce the emergence of a cyclic immune response in pathologies such as Multiple Sclerosis \cite{Compston}, \cite{Vollmer}. In this way, changing the parameters one can observe two different kind of immune response, one stationary and another cyclic. Both patologies are observed in the Multiple Sclerosis, perhaps because genetic variations between individuals turn on the immune response within a different set of parameters. To our knowledge, no other mathematical model based on the dynamics of T cells regulated by R cells had predicted it together with effective response and tolerance.

\begin{figure*}[!!htb]
    \begin{center}
        \begin{tabular}{c c c} 
           \includegraphics[angle=0, scale=0.3]{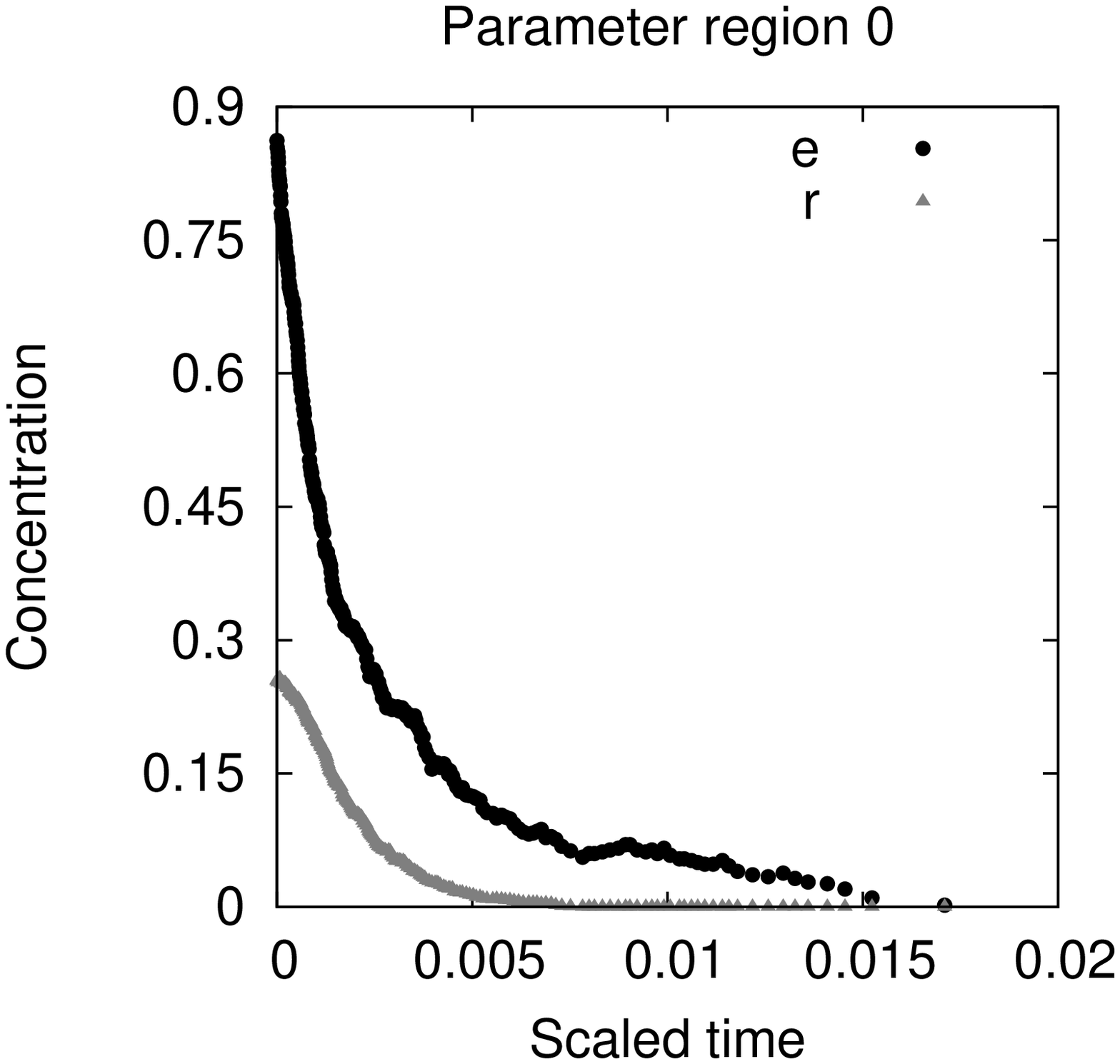}&
           \includegraphics[angle=0, scale=0.3]{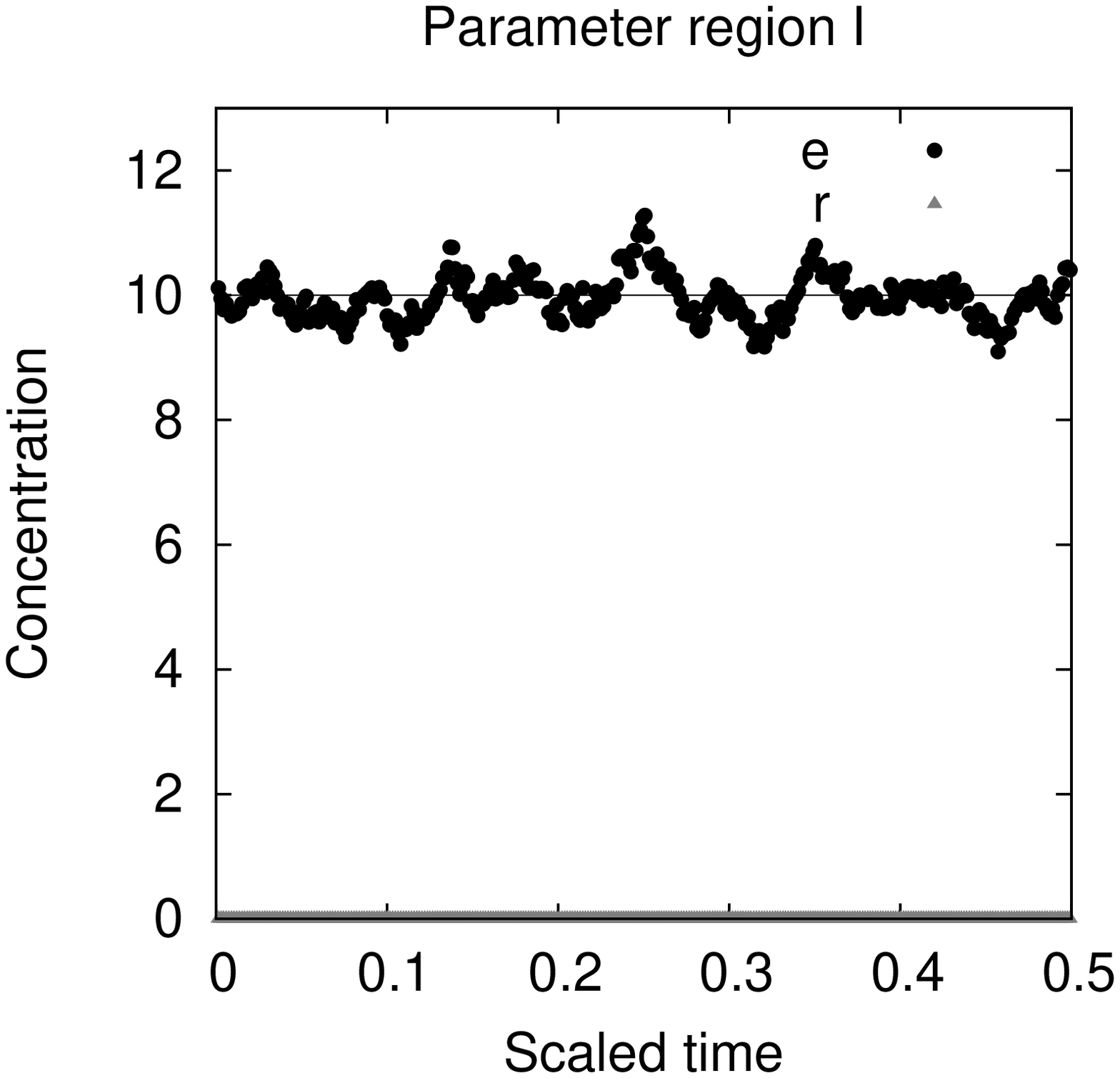}&
           \includegraphics[angle=0, scale=0.3]{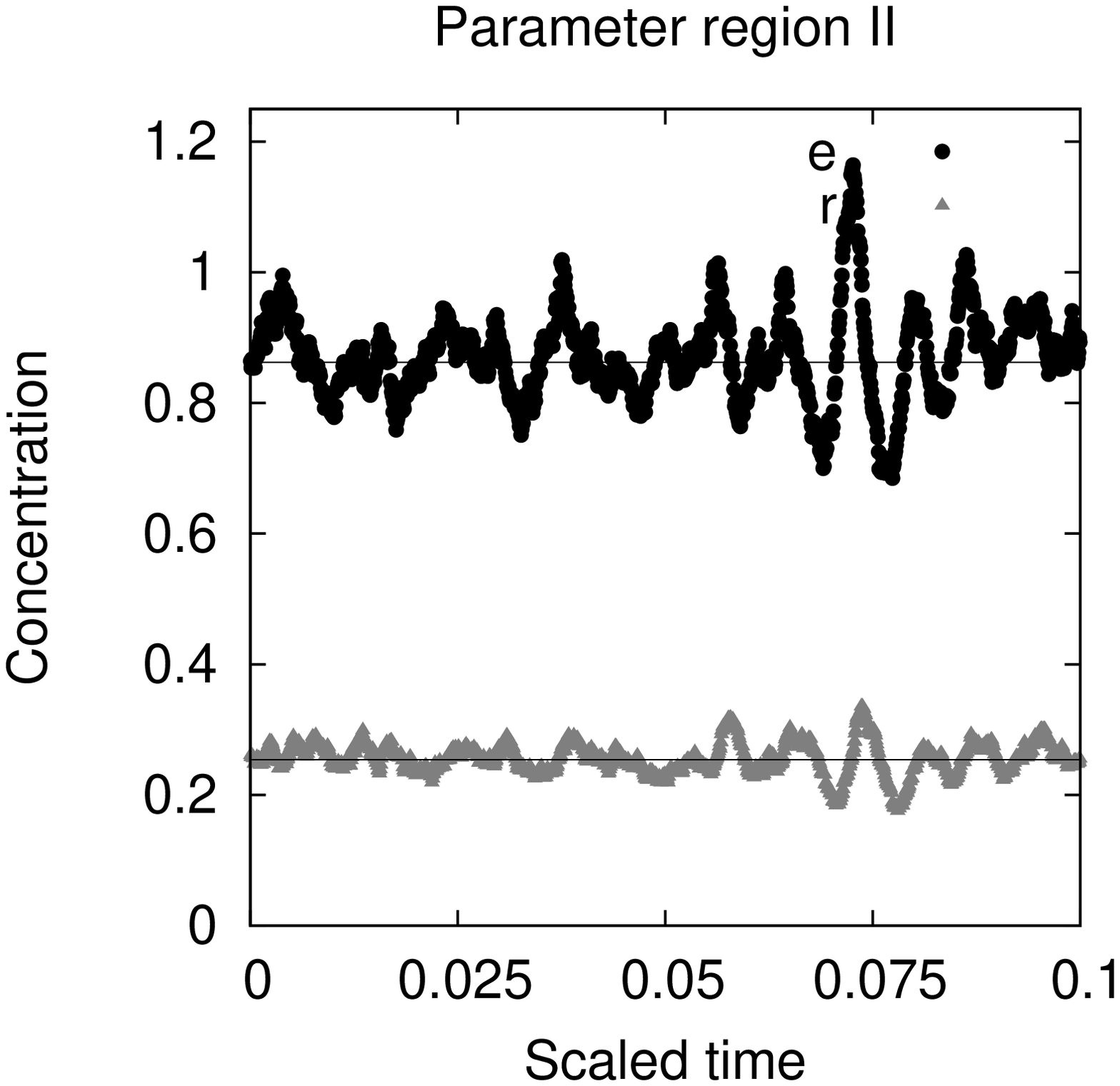}
          \\a & b & c
        \end{tabular} 
    \end{center}
\caption{ Keeping constant $s=3$, $a=1000$, $\Omega_e=1000$, $\Omega_r=2000$, $q_e=10^{-5}$, $q_r=5\cdot10^{-4}$, $m_e=m_r=1$, the figure shows results of simulations using the Gillespie's algorithm, for pairs of 
parameters [$\psi_r;\psi_e$] corresponding to different regions of the phase diaagram.
Panel \textbf{a}:[$\psi_r=185.6$ ; $\psi_e=78.0$], in region $O$, the extinction of both populations is verified.
Panel \textbf{b}:[$\psi_r=4.0$ ; $\psi_e=110.0$] in region $I$, just the effector population survives.
Panel \textbf{c}:[$\psi_r=185.6$ ; $\psi_e=198.0$] in region $II$, both species, effector and regulatory, persist.
Solid lines represent the deterministic values for fixed points, around which concentrations fluctuate. }
 \label{fig:simulaciones}
\end{figure*}

\begin{figure*}[!!htb]
     \begin{center}
             \begin{tabular}{c c c}
                 \includegraphics[angle=0, scale=0.3]{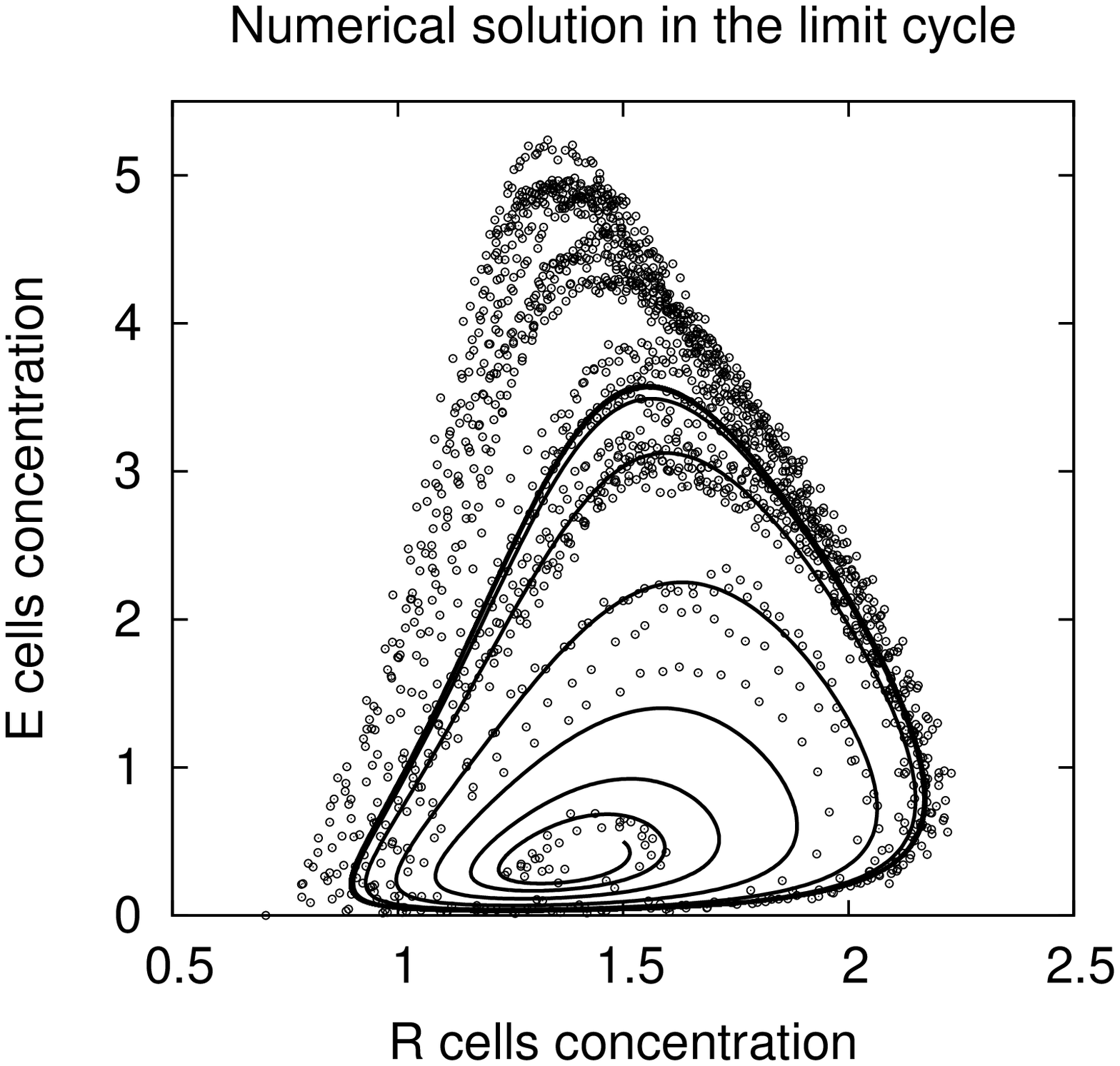} &
                 \includegraphics[angle=0, scale=0.3]{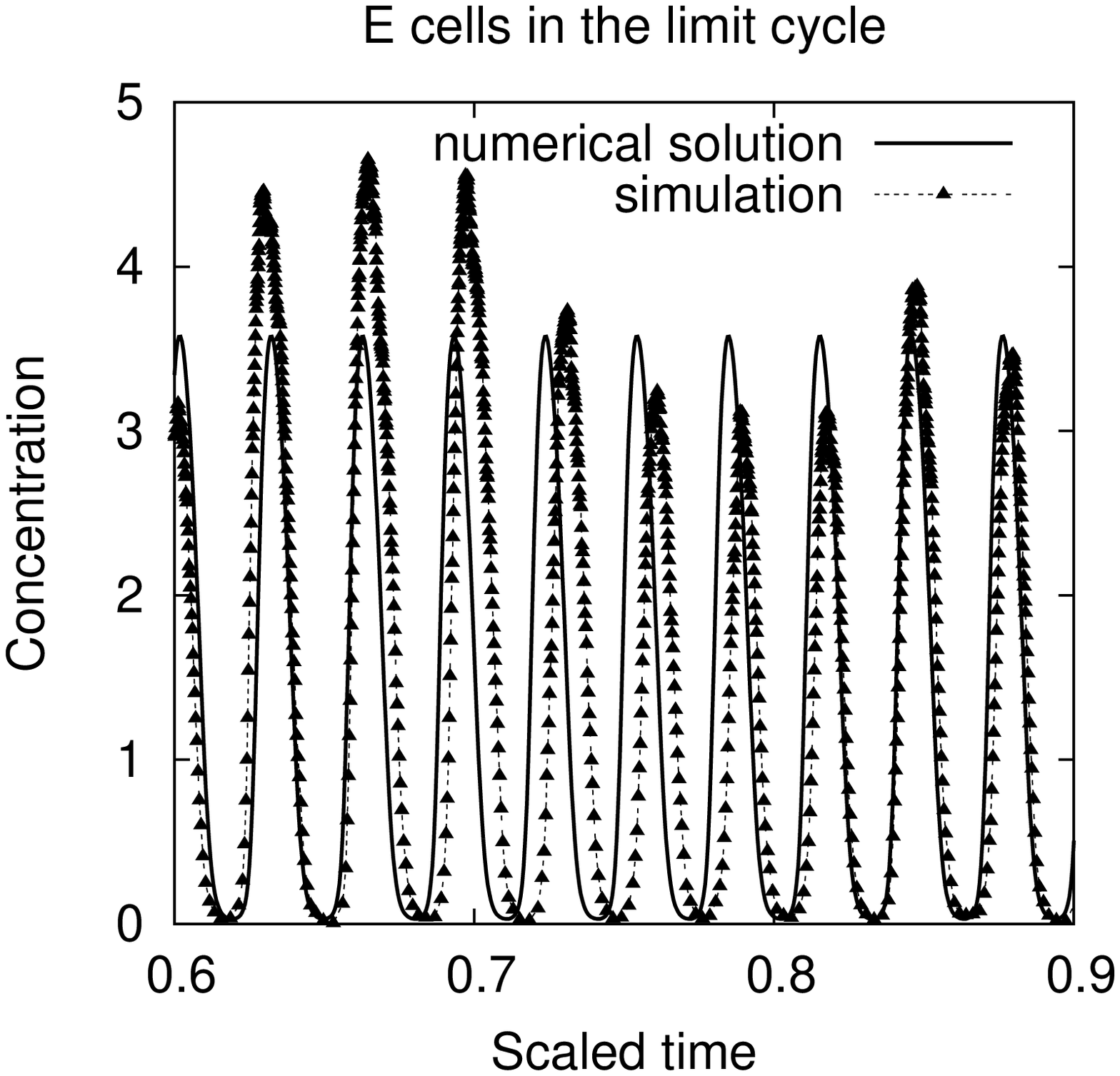} &
                 \includegraphics[angle=0, scale=0.3]{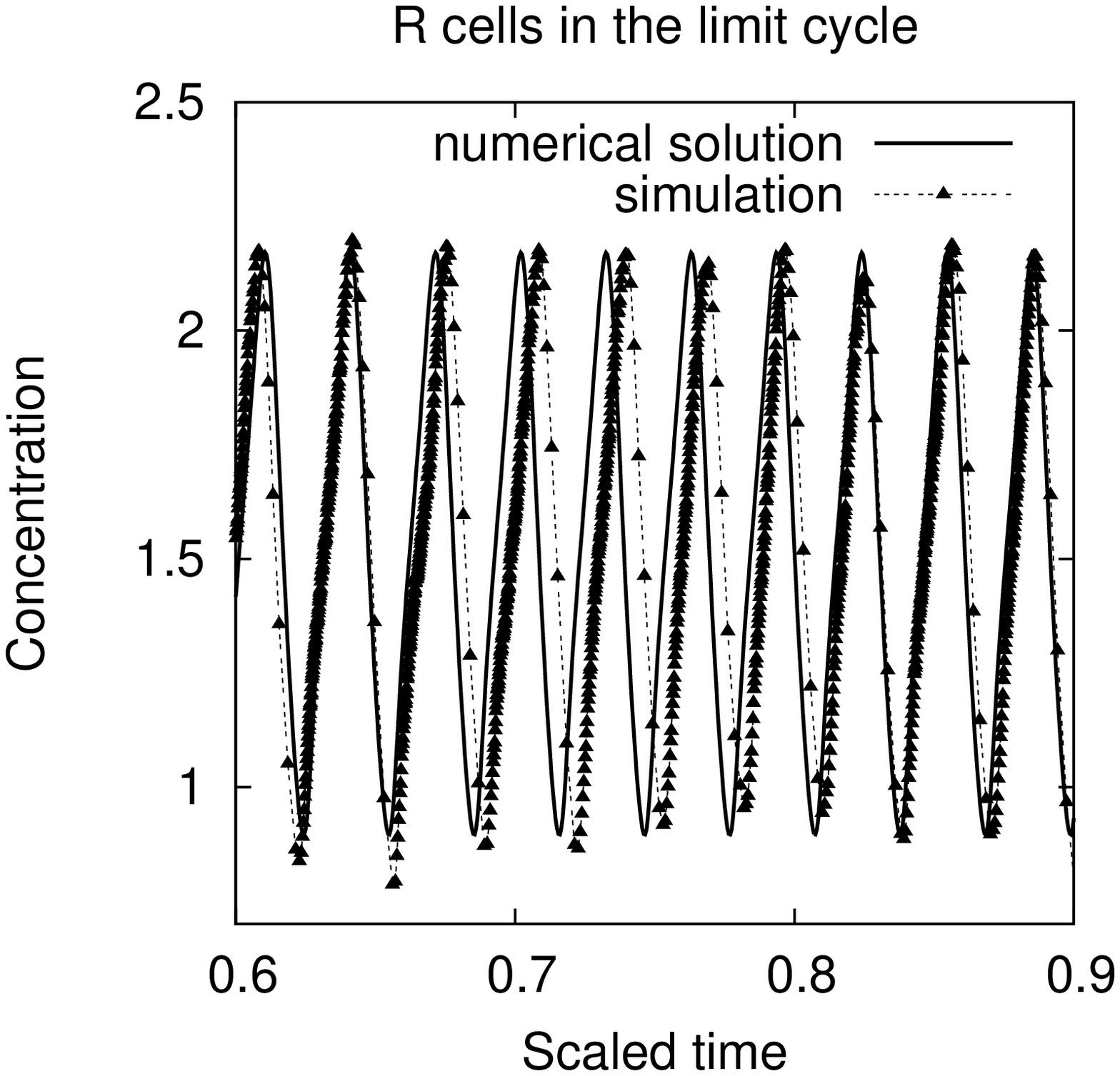}
                  \\a & b & c
             \end{tabular}
     \end{center}

 \caption{With the same parameters of Figure \ref{fig:mapa} and the pair $[\psi_r=2.2$; $\psi_e=11.0]$ from
region $IV$, the numerical solution for the deterministic rate equations (\ref{det1}) and (\ref{det2}) is compared
with a simulation made with the Gillespie's algorithm.
The trajectory surrounds the unstable fixed point ($e^{\ast},r^{\ast}$) and evolves to a cyclic behavior.
Both the amplitude and period of cycles fluctuate in the simulations. }
  \label{fig:ciclo-y-simulaciones}
\end{figure*}

Moreover, it is useful to compare the previous mean-field results with our original stochastic model.
 The reasons are two-fold, first it should confirm that the mean-field approach is correct 
when studying large systems and second, it should shed some light on the behavior of smaller systems, like those characteristics of specific responses. 
To do this, we run Gillespie's algorithm. In  Figure \ref{fig:simulaciones} we show the results of typical runs for parameter values near the different fixed points, while in Figure \ref{fig:ciclo-y-simulaciones} the numerical solutions of the deterministic equations for 
$e(\tau)$ and $r(\tau)$  are compared with the Gillespie's simulations for the cyclic regime. These graphs demonstrate that, within a stochastic approach, 
strong fluctuations appear around  the values predicted by the mean-field equations. Below, we show that if the size of the population is small enough these fluctuations become relevant and may qualitatively change the results of the mean field approximation.

\subsection{The role of the  fluctuations}

\begin{figure}[!bht]
     \begin{center}
                 \includegraphics[angle=0, scale=0.4]{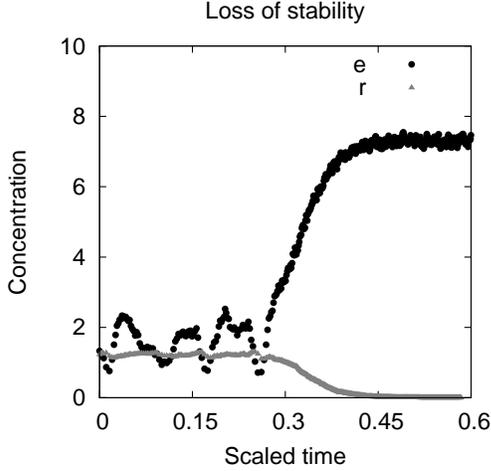}
     \end{center}
 \caption{With the fixed parameters of Figures \ref{fig:mapa} and \ref{fig:mapa-corregido}, is shown a simulation
using the pair of parameters [$\psi_r=1.17$ ; $\psi_e=8.29$] that belong to region $III$ according to 
the normally calculated eigenvalues (Figure \ref{fig:mapa}), and belong to region $I$ according to the corrected ones (Figure \ref{fig:mapa-corregido}).
 The simulation starts on the corresponding fixed point ($e^{\ast},r^{\ast}$) that was supposed to be stable, 
but fluctuations have changed its stability.
}
  \label{fig:cambioestabilidad}
\end{figure}

In smaller systems the noise may induce qualitative changes in the mean
field dynamics discussed above. This fact might be particularly
relevant, since it could imply that the dynamics of small T cells clones
would be significantly different from the one described here and in
previous works \cite{Kalet}. We have carried out a
series of Gillespie’s simulations with our stochastic model using
parameters sets that belong to region III when classified according to
the mean-field results. Our results illustrated that steady state
stabilities, specially for the steady state dominated by regulatory T
cells (tolerant steady-state), could be significantly affected by
stochastic fluctuations. Figure \ref{fig:cambioestabilidad} shows how in simulations the T cell
concentrations remain oscillating around the mean-field value of the
tolerant steady-state where the simulations began. However after a while
fluctuations drive the system away into a state dominated by effector T
cells (immune steady state). In other cases simulations shows that the tolerant steady state became unstable under stochastic fluctuations but it dynamically evolve into a trivial steady state with zero effector and regulatory T cells.

\begin{figure}[!htb]
     \begin{center}
                 \includegraphics[angle=0,scale=0.4]{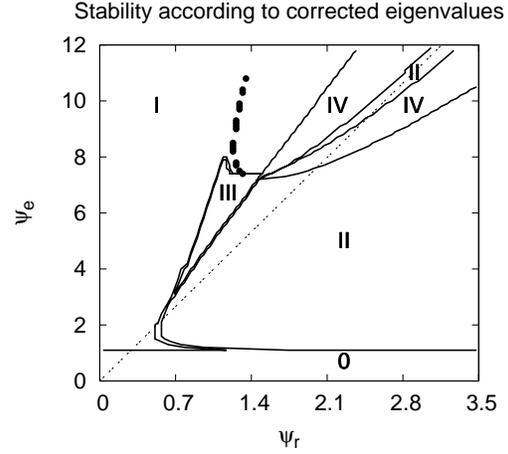}
     \end{center}
 \caption{With the same parameters of Figure \ref{fig:mapa} to build this new map, there were calculated for 
each point [$\psi_r$;$\psi_e$] the corrected eigenvalues  ${\lambda'}_i=\lambda_i + \lambda_{i_{corr}}$, 
and according to the signs of these, it was classified every point in one of the regions $O$ to $IV$.
Black points correspond to region $III$.
}
  \label{fig:mapa-corregido}
\end{figure}

A new phase diagram (predicted with the help of (\ref{lambdacorregida})
is represented in Figure \ref{fig:mapa-corregido}, and has some differences when compared with Figure \ref{fig:mapa}. The more striking feature of this map is the apparition of a new zone embedded in region $IV$ with the characteristics of phase $II$. In this region of the parameters, instead of having a cyclic behavior, the system  evolves into a tolerant state (region $II$). This conclusion can  be corroborated by simulations, and is in contradiction with what the normal analysis of eigenvalues predicts. 

This result sustains the idea that, increasing the number of APCs, a cyclic immune response may be turned into a tolerant response (see the dashed line joining regions $IV$ and $II$ in Figure \ref{fig:mapa-corregido}). Along this line,  when the number of APCs is increased, simulations show a stretching of the amplitude of the cycles until they reduce to a mere fluctuating behavior around the deterministic fixed point as in Figure \ref{fig:simulaciones} \textbf{c}.

On the other hand, fluctuations, also reduced the bistable region $III$ favoring the immune response of the system. Note however, how small spots of bistability persist within region $I$.

Moreover, we study the effect of these fluctuations when the system is subject to external perturbations. For example, in the phase III, 
the system, depending on the initial conditions may stay in the tolerant state $(e>0,r>0)$ or in the state  $(e>0,r=0)$. In the mean field model, to switch the system from one state to the other, it must be strongly perturbed and moved away from the respective basin of attraction of the fixed point. In the case of small systems, small perturbations may switch the state of the system, provided they are frequent enough. 

\begin{figure}[!htb]
     \begin{center}
                 \includegraphics[angle=0, scale=0.6]{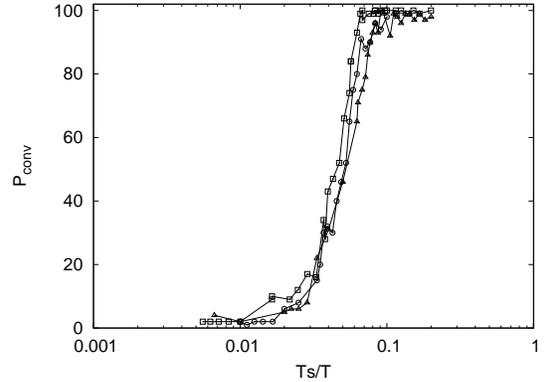}
     \end{center}
 \caption{Transition from the tolerant state $(e>0,r>0)$ to the responsive state $(e>0,r=0)$ as a function of the number of external perturbations when the system is in region $III$. 
}
  \label{fig:mypert}
\end{figure}

In figure \ref{fig:mypert} we show the probability ($P_c$) of switching from the tolerant state to the state $(e>0,r=0)$ as a function of the number of perturbations. In this case, with a period $T$ the number of cells at a given time was reduced an 80 percent. $T_s$ is the size of the simulations (different symbols mean three different $T_s$). As can be easily see if the number of perturbations is small enough $T_s/T \rightarrow 0$, the system do not switch from one state to the other. It is always tolerant. On the other hand, when $T_s/T \rightarrow \infty$, i.e. when the period of the perturbations is small enough,  the system goes to the responsive state with probability 1. The existence of a clear threshold for the minimum number of perturbations needed to change the state also signatures the importance of the fluctuations for the occurrence of this transition. 

All these results confirm our previous intuition. The role of the fluctuations may be relevant and may alter significantly the predictions of the mean-field models. In particular, regions that were considered bi-stable may now develop immune response, and regions in which a cyclic pattern was predicted may become fully tolerant. Moreover, also the response of the system to external perturbations change. Perturbations, that may be produce only transient behaviours in the deterministic model, may become relevant when fluctuations are taken into account.
Of course, the magnitude of these results are parameter dependent and probably richer diagrams may be found exploring further the model. But our results already prove that the interpretation of the specific immune response in base of mean field models should be done with care.

\subsection{Power spectrum}

Any single run of the stochastic system (as panels \textbf{b} and \textbf{c} in Figure \ref{fig:simulaciones}) shows that coherent oscillations are sustained in parameter regimes in which the 
deterministic equations approach a fixed point. It evidences the fact that fluctuations can not be safely ignored in 
systems composed of a few thousands of elements. 

Therefore, it is important to understand, not only the consequences of these fluctuations as done above, but also their properties. To do this, we 
analyze the characteristics of these fluctuations around the steady state, using 
equation (\ref{espectropotencias1}): 

\begin{equation}
P_i(\omega) =  \sum_{j=1}^N \sum_{k=1}^N \Phi^{-1}_{ij}(\omega) D_{jk} (\Phi^\dag)^{-1}_{ki}(\omega) \mbox{,}
\label{espectropotencias2}
\end{equation}
 keeping in mind that for our model

\[\textbf{A}= \left( \begin{array}{*{2}{c}}
\frac{\partial W_1}{\partial e}-\frac{\partial W_3}{\partial e} & \frac{\partial W_1}{\partial r}-\frac{\partial W_3}{\partial r}  \\
  &  \\
\frac{\Omega_e}{\Omega_r}\left(\frac{\partial W_2}{\partial r}-\frac{\partial W_4}{\partial r}\right)  & \frac{\Omega_e}{\Omega_r} \left( \frac{\partial W_2}{\partial r}-\frac{\partial W_4}{\partial r}\right)  

\end{array} \right) \]
and

\[\textbf{D}= \left( \begin{array}{*{2}{c}}
 W_1+ W_3 & 0  \\
  &  \\
 0 & \left( \frac{\Omega_e}{\Omega_r}\right) ^2 \left( W_2+ W_4\right)  
\end{array}   \right)\mbox{.} \]
Then, evaluating $\textbf{A}$ and $\textbf{D}$ in the corresponding fixed point, and using them to calculate eq. 
(\ref{espectropotencias2}) we can determine the contribution of every frequency to the composition of 
the oscillations. Then, the discrete Fourier transform of a data of concentration and time that comes from the simulations can be compared with the power spectra curve calculated with (\ref{espectropotencias2}). It would give us a way to know how good the made approximations were. 

\begin{figure}[!htb]
     \begin{center}
                 \includegraphics[angle=0, scale=0.4]{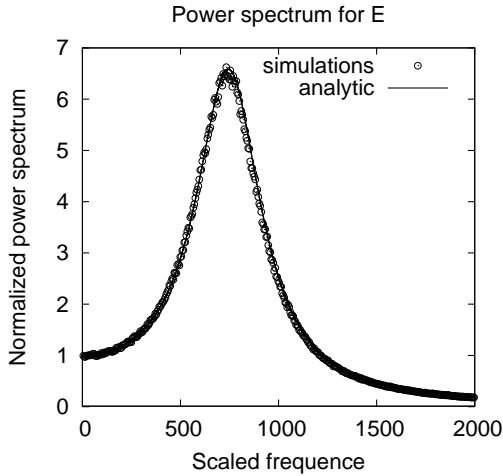}
     \end{center}
 \caption{Superposition of the power spectrum for E cells calculated with the analytic expression
 (\ref{espectropotencias2}) and the square of the absolute value of the Fourier transform of 6000 Gillespie's
simulations. The normalization condition is $P(\omega=0)=1$.
The used parameters are [$\psi_r=185.6$ ; $\psi_e=198.0$]; $q_e=10^{-5}$; $q_r=5\cdot10^{-4}$; $\Omega_1=1000$; 
$\Omega_2=2000$; $m_e=m_r=1$; $s=3$ and $\alpha=1000$, corresponding to region $II$.
}
  \label{fig:espectroE}
\end{figure}

In Figure \ref{fig:espectroE} we show the power spectra calculated by (\ref{espectropotencias2}) and the one obtained by averaging the Fourier transforms of 6000 different Gillespie's simulations.
Both have been normalized in a way that $P(\omega=0)=1$ and it is evident that the coincidence is remarkably good.

The peak in the power spectrum indicates the existence of a resonance frequency. In it, the fluctuations
of the concentrations that define the state of the system amplify their values as a function of the
 parameters of the model. This effect may be interpreted as the induction by the internal noise of
a a kind of stochastic resonance that  is responsible for the emerging of amplified  oscillations in systems which, when described through the deterministic rate equations for a given set of parameters lack a cyclic behavior.

From the technical point of view, this characteristic frequency arises because of the existence of complex eigenvalues with non-zero imaginary parts for the Jacobian at the fixed point. In the standard stability analysis, the imaginary part of the eigenvalues are responsible for oscillatory transient behaviors near the fixed points. However, here the stochasticity due to the smallness of the system
 results in persistent perturbations away from this fixed point, so that both features together result in an overall oscillatory effect.

The existence of this characteristic frequency could be useful to induce transition between two states of the 
system just by making a parameter (such like the number of APCs) to fluctuate with a small amplitude but with the
appropriate frequency. Simulations corroborate this when it is induced a change from a tolerant phase to an immune
one by making $a$ to fluctuate near a $III$-$I$ region border. Work is in progress to check the reliability of this approach to develop new therapeutic techniques.

\section{Conclusions}
\label{sec:conc}

We present a stochastic model of  interactions between Effector and Regulatory CD4+ T cells in the presence of Antigen Presenting Cells that is able to show  active, tolerance, or cyclic immune responses. We characterize the phase diagram in the mean field approximation, and then analyze the corrections, due to finite size effects, to this phase diagram in the presence of fluctuations. This corrected phase diagram may be understood as a first attempt to characterize a stochastic model for the specific response of the immune system. In fact, we prove that the fluctuations may strongly alter the mean field predictions turning for example tolerant zones in responsive or cyclic responses into tolerant, and therefore that any analysis of the clonal response of the immune system based on mean field predictions must be taken with caution. Moreover, we show that our model present sustained oscillations at a characteristic frequency  that  may be relevant to understand, or treat these clonal responses.


\section*{Appendix A: Gillespie's Algorithm}

The occurrence of every process that can take place into our system has a very strong stochastic component. Every
reaction depends on a lot of variables that can only be described in a probabilistic way, so in a given state,
 both the next process that would take place, and the moment in which it will occur are random variables. Even when
it is not  possible to find an analytic solution for the Master equation that governs the system, microscopic 
simulations of the processes defined by the relations (\ref{W1}) to (\ref{W4}) can be carried out using the algorithm originally proposed by Gillespie \cite{Gillespie}.

The algorithm keeps the random fashion in which processes occur. The backbone of such a simulation consists in 
determining which one will be the next process, and when will it occur. Every run of the simulation will give 
a different temporal sequence of concentration that is, according with the Master equation of the system. 

The procedure can be synthesized in three main steps:

\begin{enumerate}
 \item
Calculate the occurrence probabilities ($W_j$) per unit time for every process that can take place in the system.
 \item 
Generate two random numbers $r_1$ and $r_2$ using a unit-interval uniform random number generator, and calculate 
$\Delta t$ and $\mu$ according to (\ref{tau}) and (\ref{mu}):
\begin{equation}
   \Delta t = \frac{1}{W} \ln\left( \frac{1}{r_1} \right) \;\;\;\;\ \mbox{and}
\label{tau}
\end{equation} 
\begin{equation}
  \sum_{j=1}^{\mu-1} W_{j} < r_2 W\leq  \sum_{j=1}^{\mu} W_{j} \mbox{,}
\label{mu}
\end{equation} 
where 
\begin{equation}
 W=\sum_{j=1}^M W_j \mbox{.}
\end{equation} 
 \item 
Using the $\Delta t$ and $\mu$ values obtained, increase the time $t$ by $\Delta t$, and adjust the cellular population levels
to reflect the occurrence of the $\mu^{th}$ process. 
\end{enumerate}

\bibliographystyle{plain}
\bibliography{bibliografia}

\end{document}